\documentclass[12pt,preprint]{aastex}

\slugcomment {}

\shorttitle{The nearby, young, isolated, dusty star HD 161991}
\shortauthors{Schneider et al.}

\begin{document}

\title{The nearby, young, isolated, dusty star HD 166191}

\author{Adam Schneider}
\affil{Department of Physics and Astronomy, University of Georgia,
    Athens, GA 30602}
\email{aschneid@physast.uga.edu}

\author{Inseok Song}
\affil{Department of Physics and Astronomy, University of Georgia,
    Athens, GA 30602}
\email{song@physast.uga.edu}

\author{Carl Melis\footnote{NSF Fellow}}
\affil{Center for Astrophysics and Space Sciences, University of California,
    San Diego, CA 92093}
\email{cmelis@ucsd.edu}

\author{B. Zuckerman}
\affil{Department of Physics and Astronomy, University of California,
    Los Angeles, CA, 90095}
\email{ben@astro.ucla.edu}

\author{Mike Bessell}
\affil{Research School of Astronomy and Astrophysics, The Australian National University,
    Weston Creek, ACT 2611, Australia}
\email{bessell@mso.anu.edu.au}

\author{Tara Hufford}
\affil{Department of Physics and Astronomy, University of Georgia,
    Athens, GA 30602}
\email{tara@physast.uga.edu}

\and

\author{Sasha Hinkley$^1$}
\affil{Department of Astronomy, California Institute of Technology,
	Pasadena, CA 91125}
\email{shinkley@astro.caltech.edu}

\begin{abstract}
\end{abstract}

We report an in-depth study of the F8-type star HD 166191, identified in an ongoing survey for stars exhibiting infrared emission above their expected photospheres in the Wide-field Infrared Survey Explorer all-sky catalog.  The fractional IR luminosity measured from 3.5 to 70 $\mu$m is exceptionally high (L$_{IR}$/L$_{bol}$ $\sim$10\%). Near-diffraction limited imaging observations with the T-ReCS Si filter set on the Gemini South telescope and adaptive optics imaging with the NIRC2 Lp filter on the Keck II telescope confirmed that the excess emission coincides with the star.  Si-band images show a strong solid-state emission feature at $\sim$10 $\mu$m.  Theoretical evolutionary isochrones and optical spectroscopic observations indicate a stellar age in the range 10-100 Myr.  The large dust mass seen in HD 166191's terrestrial planet zone is indicative of a recent collision between planetary embryos or massive ongoing collisional grinding associated with planet building.


\section{Introduction}

The vast majority of debris disks discovered to date are cold (T$_{dust}$ $\leq$ 150K) and exist at distances of ten to hundreds of AU from their host star, many of which were discovered by the {\it Infrared Astronomical Satellite (IRAS)} (Chen et al.\ 2005, 2006; Bryden et al.\ 2006, 2009; \citealt{rhee07}).  Instances of warm dust (T $\geq$ 150K) are much rarer (\citealt{song05}, \citealt{cur07}, \citealt{rhee08}, \citealt{smith08}, \citealt{melis10}, and \citealt{mor11}).  A rapid rise of dust luminosity at young stellar ages due to warm dust is predicted by simulations, and is ascribed to the beginning stages of oligarchic growth \citep{ken08}.  Currie et al. (2008a, 2008b, 2009) find a dust luminosity peak around 10-20 Myr for intermediate mass stars.  \cite{melis10} find the epoch of final mass accretion of terrestrial planets to be 30-100 Myr for solar type stars, and 10-30 Myr for intermediate mass stars (1.5-8.0 M$_{\sun}$).  Many of these warm dust systems have been found to have multiple components in the radial direction (\citealt{mor11} and references therein).  But when the warm dust is the dominant component in the debris disk, then oftentimes there is no evidence for cooler dust \citep{melis10}.

The Wide-field Infrared Survey Explorer (WISE) all-sky survey presents an unprecedented opportunity to explore nearby stars for excess emission at mid-infrared wavelengths indicative of warm dusty circumstellar disks.  Because circumstellar disks are ultimately linked to a star's final planetary architecture, their study can provide insight into how planets form and are useful as targets in extrasolar planet searches.  We present here an observational study and SED analysis of the dust surrounding the F8-type star HD166191.

\section{Identification of HD 166191}
As part of an ongoing effort to identify stars with IR-excess we have cross-correlated the WISE all-sky survey catalog with the Tycho 2 optical star catalog \citep{hog00}.  In an attempt to limit the search to nearby stars, we adopted a total minimum proper motion threshold of 25 mas yr$^{-1}$.  For each Tycho star, we fit a synthetic stellar spectrum to the visual (Tycho-2) and near-IR (2MASS) photometric data using a $\chi$$^2$ minimization algorithm following the methods outlined in \cite{rhee07}.  Output from this SED-fitting algorithm yields a significance of excess parameter for each filter band, defined as (F$_{meas}$ - F$_{phot}$) / $\sigma$$_{F_{meas}}$, where F$_{meas}$ is the measured flux, F$_{phot}$ is the estimated photospheric flux for the primary stellar type, and $\sigma$$_{F_{meas}}$ is the associated measurement uncertainty.  Our significance of excess evaluation only takes into account the WISE photometric uncertainties and not any uncertainty from the spectral fitting.  Our first inspection, aimed at identifying the largest excess sources, specified a significance of excess greater than fifty in both WISE channel 3 and 4 (W3 and W4, respectively).  This search returned 44 total stars (out of a total sample of $\sim$500,000), the majority of which fell into one of three categories: pre-main sequence stars (17 total - e.g. HD 282624, HD 142560, and HD 36910), Be stars (10 total - e.g. HD 22191, HR 3858, and HD 105435), and well-studied debris disk hosting stars (7 total - e.g $\beta$ Pic, TW Hya, and BD+20 307).  There are ten potentially new excess sources that need additional confirmation and characterization.  Within this sample though, is HD 166191 (aka HIP 89046, TYC 6843-1557-1) for which there is relatively little information regarding its circumstellar dust.  Further inspection of HD 166191 revealed excess at 3.6 and 4.5 $\mu$m as well.  This star was notable amongst other IR-excess candidates because of its large apparent excess and relative obscurity in the literature; this star has largely been ignored by previous searches for circumstellar disks.  Due to the little attention paid to this star up to now, we initiated a more in depth study of the origin and nature of HD 166191 and it's circumstellar dust.

\section{Observations}

\subsection{Previous Photometry}

After identifying HD 166191 as a WISE excess source, we queried additional  catalogs and archives for supplementary photometric measurements.  This star was detected at 12, 25, and 60 $\mu$m, with an upper limit at 100 $\mu$m in the IRAS Point Source Catalog \citep{beich88}. Inspection of the 60 $\mu$m detection utilizing the {\it IRAS} Scan Processing and Integration (Scanpi) software indicates substantial contamination by nearby IR cirrus, which likely dominates the 100 $\mu$m upper limit as well.  The {\it IRAS} Faint Source Catalog avoided the galactic plane and does not contain HD 166191 (b = -2 deg).  HD 166191 was also detected by the Midcourse Space Experiment ({\it MSX}; \citealt{price01}) at 8.3, 12.1, 14.7, and 21.3 $\mu$m and the AKARI satellite \citep{ish10} at 18 $\mu$m. According to \cite{fuj13}, HD 166191 was observed at 9 $\mu$m during a South Atlantic Anomaly passage, and hence rejected from the AKARI PSC.  12 and 25 $\mu$m flux densities from {\it IRAS} and at five wavelengths from {\it MSX} and AKARI are provided in Table 1.

We also queried the Spitzer Heritage Archive to examine if HD 166191 was observed by any of Spitzer's instruments.  The star was observed with the Infrared Array Camera (IRAC; \citealt{faz04}) as part of the Galactic Legacy Infrared Mid-Plane Survey Extraordinaire (GLIMPSE; \citealt{ben03}) at 3.6, 4.5, 5.8, and 8.0 $\mu$m.  HD 166191 was also observed with the Multiband Imaging Photometer for Spitzer (MIPS; \citealt{rieke04}) at 24, and 70 $\mu$m as part of the MIPSGAL survey \citep{car09}, a MIPS complimentary survey to GLIMPSE.  We constructed a mosaic image using the SSC mosaicking and point-source extraction (MOPEX) software \citep{mak05} at each MIPS wavelength.  A standard aperture photometry procedure (aperture radius of 5 pixels and sky annulus of 10-20 pixels) was utilized to extract source counts at each wavelength.  IRAC flux densities from the GLIMPSE catalog and calculated MIPS flux densities are listed in Table 1.  All fluxes listed in Table 1 are not color-corrected, though they are color-corrected for the creation of Figure 1 (see Section 4.1).

\subsection{Infrared Photometry and Optical Spectroscopy}

We observed HD 166191 on 3 October 2012 UT with an entire set of narrow Silicate filters with T-ReCS \citep{tel98} at Gemini South (Gemini observation GS-2012B-Q-90-284, Si filter names and central wavelengths given in Table 1).  We used the superior mid-IR angular resolution of Gemini South ($\sim$0.4\arcsec) compared with existing images ($>$ 2\arcsec), to rule out the presence of an unresolved source of mid-IR contamination (as was found for the reddest Aumann catalogue IRAS source by \cite{lisse02}), and also to inspect for possible solid state features.  The T-ReCS images of HD 166191 were approximately seeing limited and show no evidence of additional sources at any wavelength in the $\sim$20$\arcsec$ $\times$ 30$\arcsec$ field of view.  For the single detected source at the expected position of HD 166191, we measure a FWHM of $\sim$ 0\farcs5 in the Si-1 filter image.  The measured FWHM is fully consistent with a point source under such observing conditions.

With the Keck II telescope, we obtained adaptive optics images of HD 166191 with the NIRC2 camera J, H, K$_S$ filters on 1 May 2013 UT and the 3.8 $\mu$m (Lp) filter on 31 May 2013 UT.  All images were taken with the narrow-field camera (pixel scale = 10 mas pixel$^{-1}$). The noisy lower left quadrant was avoided by employing a three position dither pattern.  Each J, H, and K$_S$ frame had an exposure time of 1.81 seconds (0.181 second single frame integration times $\times$ 10 coadds), and 6 separate exposures, resulting in total integration times of 10.86 seconds. The Lp frames had exposure times of 5.3 seconds (0.053 second single frame integration times $\times$ 100 coadds) and 24 separate exposures, resulting in a total integration time of 127.2 seconds.  The data were reduced using custom scripts to subtract the dark current, apply a flat-field correction, shift, and median combine to create final images.  The aim of these observations was to clearly show that HD 166191 is unresolved into multiple components and further explore the possible presence of contaminating objects.  The results of these observations are discussed in section 4.2.                 

To obtain age diagnostics from an optical spectrum, we observed HD 166191 with the Wide Field Spectrograph (WiFeS) and echelle spectrograph at the Siding Spring Observatory's 2.3 m telescope on 2012 November 03 (UT) and 2012 November 28 (UT), respectively.  The measured Li $\lambda$6708 equivalent width (which is consistent in the WiFeS and echelle spectra), as well as the echelle measured radial velocity, are given in Table 2 and discussed in section 4.3.

\section{Results}

\subsection{Spectral Energy Distribution}

The spectral energy distribution of HD 166191 as derived from the Table 1 photometry is shown in Figure 1.  Every measured flux beyond 3 $\mu$m, which includes measurements from WISE, IRAS, IRAC, MIPS, MSX, AKARI, and T-ReCS (23 measurements in total) lies well above the level expected from a pure stellar photosphere, indicating substantial amounts of circumstellar dust.  The shape of the excess emission is complex, and is not easily modeled by a simple blackbody fit, contrary to most known IR-excess sources.  While not necessarily a complete or true representation of the system, a simple disk model can aid in estimating dust properties.  The excess emission is fit using two blackbodies with temperatures of 760 and 175 K. The solid state dust emission feature around $\sim$10 $\mu$m influences the photometry from at least four different instruments, including WISE, IRAC, MSX, and T-ReCS and indicates copious amounts of sub-micron to micron sized warm dust grains.  Color corrections are applied to all broadband photometric measurements at wavelengths greater than 3 $\mu$m by determining the power law slope at each filter band using the combined stellar and blackbody dust fit and applying the relations found in \cite{wri10}, the IRC Data User's Manual\footnote{http://www.sciops.esa.int/SA/ASTROF/docs/IRC\_IDUM\_1.3.pdf}, \cite{beich88}, the IRAC Instrument Handbook\footnote{http://irsa.ipac.caltech.edu/data/SPITZER/docs/irac/iracinstrumenthandbook/}, the MIPS Instrument Handbook\footnote{http://irsa.ipac.caltech.edu/data/SPITZER/docs/mips/mipsinstrumenthandbook/}, and the Midcourse Space Experiment Point Source Catalog Version 1.2 Explanatory Guide\footnote{http://irsa.ipac.caltech.edu/data/MSX/docs/MSX\_psc\_es.pdf}.

We note here the discrepancy between the measured Spitzer IRAC flux and WISE flux at $\sim$4.5 $\mu$m. We speculate that this difference is likely due to the higher resolution of the IRAC instrument compared to that of WISE.  There is also the possibility that this difference is evidence of temporal variability in the flux at this wavelength, not unlike the variability seen in \cite{meng12} and the extreme case of TYC 8241-2652-1 \citep{melis12}.     

\subsection{The age of HD 166191}

The age of HD 166191 can give us clues as to whether the dusty disk is primitive or a true debris disk.  The age of HD 166191 was estimated following  methods in \cite{zuck04}.  We first inspect the position of HD 166191 on a color-magnitude diagram (CMD) and find that it lies above the zero-age main-sequence, consistent with a 3$\sigma$ age range between $\sim$2 and 15 Myr via comparison with theoretical isochrones from \cite{bre12} (Figure 2).  Because the presence of an unresolved companion can significantly influence CMD position, leading to an inaccurate age estimate, we investigated for any possible hint of binarity.  No indication of double lined binarity is present in the single epoch echelle spectrum of HD 166191. We also sought out any potential companions in available images.  The highest resolution publicly available image for HD 166191 in the optical or near-IR comes from the UKIRT Infrared Deep Sky Survey (UKIDSS; \citealt{law07}) Galactic Plane Survey (GPS; \citealt{luc08}).  This image does reveal the presence of a possible companion $\sim$3$\arcsec$ away that is unresolved in 2MASS, though with a $\Delta$K of $\sim$4.3, we do not believe this source is bright enough to significantly affect HD 166191's photometry.  This source is not present in any of the T-ReCS images because they were not sensitive enough to detect a non-dusty faint star.  Keck NIRC2 Lp images reveal the presence of two nearby objects, both approximately 1$\arcsec$ away (Figure 3).  The measured flux ratios of these nearby objects to HD 166191 are $\sim$4.1\% and $\sim$0.2\% for `1' and `2', respectively.    

To evaluate the accuracy of age estimation utilizing theoretical isochrones on a CMD, we compare HD 166191's CMD position with members of the Beta Pictoris Moving Group (BPMG) and the Upper Centaurus Lupus (UCL) region of the Scorpius Centaurus OB association (Figure 2).  The BPMG was chosen because it has a well constrained age ($\sim$12 Myr) derived from a kinematic traceback analysis \citep{ort02}.  The UCL region was chosen because of the possibility that HD166191 is an outlying member (see Section 5.1).  Only accepted members of UCL with well-determined parallaxes from \cite{chen11} were chosen for the figure.  While the non-binary members of the BPMG coincide reasonably well with the $\sim$12 Myr isochrone, the figure indicates that, even for confirmed UCL members with no known binarity, there is significant scatter due to large distance uncertainties.  Considering the possibility of being an unresolved multiple system, we very conservatively estimate the CMD age of HD 166191 to be 5$^{+25}_{-3}$ Myr.

Lithium content, in addition to being age dependent, is strongly correlated with stellar mass.  The measured equivalent width for HD 166191 (120 $\pm$ 5 m\AA) supports an age $\lesssim$ 100 Myr \citep{zuck04}.  We also compare the calculated UVW space motions in Table 2 to those of known nearby stellar associations and find that HD 166191 does lie within the ``good box'' defined in \cite{zuck04} for young stars (see their Figure 6).  No X-ray emission from HD 166191 is reported in the {\it ROSAT} All-Sky Survey (RASS), but, from the XMM-Newton Serendipitous Source Catalog \citep{wat09}, we calculate a log(L$_{X}$/L$_{bol}$) value of $\approx$ -4.9, indicating an age $\gtrsim$ 100 Myr (see Figure 4 of \citealt{zuck04}).

For sun-like stars, protoplanetary disk dissipation is thought typically to occur by 6 Myr (e.g., \citealt{wyatt08}), although there are exceptions (e.g., TW Hya, HD 98800, V 4046 Sgr).  If the disk around HD 166191 is in a primordial or transitional stage, one might expect several observable spectral signatures of active accretion, such as H$\alpha$ emission; HD 166191 shows no such signs.  The H$\alpha$ absorption equivalent width is given in Table 2. Instead, if HD 1666191 is a weak-line T Tauri star (WTTS) with age $\leq$ 6 Myr, we would anticipate strong X-ray emission (X-ray luminosity $>$ 10$^{30}$ erg s$^{-1}$, or log(L$_{X}$/L$_{bol}$) $>$ -4.3 for the spectral type of HD 166191 - \citealt{owen11}), but this is not the case.  Lastly, assuming that the disk is essentially gas free, we evaluate the timescales of small grain survival due to Poynting-Robertson (PR) drag and collisional grinding using equation (5) from \cite{chen01} and equation (5) from \cite{john12}, respectively.  Because HD 166191 has a very large fractional infrared luminosity due to the warm dust component ($\sim$6\%, see Section 5.2), the collisional timescale is very short - less than a few years for a range of orbital distances (0.3 - 1.0 AU).  Assuming a grain density of 2.5 g cm$^{-3}$ and a range of grains sizes (0.1-10 $\mu$m) and orbital distances (0.3 - 1.0 AU), the range of particle lifetimes due to PR drag is $\sim$100 - 8000 yr.  These timescales are much smaller than the age of HD 166191, indicating that, if the disk is gas-free, any primordial dust should have been cleared from it within 10,000 yrs.  Thus, any currently present dust must have been created from secondary sources, and then it is not primordial in nature.  

Most indicators for HD 166191 are suggestive of a young age.  We assign a conservative estimate of 10-100 Myr.

\section{Discussion}

\subsection{Origin of HD 166191}

HD 166191 is too distant (d $\sim$119 pc) to belong to any of the nearby young moving groups of comparable age (AB Dor, $\beta$ pic, Tuc-Hor, etc...).  However, it's distance is similar to that of the Scorpius Centaurus association (ScoCen), and it has a reasonable separation from the historically defined ScoCen region to be considered a potential outlying member.  It is $\sim$7$\degr$ from the nearest edge defining the Upper Centaurus Lupus subregion (based on \citealt{dez99}).  To determine if HD 166191 is a possible UCL outlier, we checked to see if it has UVW space motions consistent with the UCL subgroup. First, we cross-matched the suggested Hipparcos members from \cite{dez99} with the Extended Hipparcos Compilation (XHIP; \citealt{and12}) to obtain radial velocity measurements, retaining only those members with a radial velocity quality flag of A or B ($\sim$200 total stars).  Using robust statistics \citep{hoag83} to exclude possible non-members, we calculated an average UVW for UCL of (U = -8.4 $\pm$ 2.7, V = -19.1 $\pm$ 3.8, and W = -5.5 $\pm$ 2.0), values consistent with those found in \cite{chen11} (U = -5.1 $\pm$ 0.6, V = -19.7 $\pm$ 0.4, and W = -4.6 $\pm$ 0.3).  The UVW space motion of HD 166191 (U = -5.6 $\pm$ 1.3, V = -21.9 $\pm$ 3.0, and W = -8.7 $\pm$ 1.5) is consistent with UCL space motions, so while this star resides $\sim$7$\degr$ from the edge of the UCL boundary defined in \cite{dez99}, there is a possible connection.

HD 166191 is located in an extremely crowded field near the galactic center, therefore a companion search based solely on proximity or common proper motion is impractical.  For this reason, we searched the area around HD 166191 for any stars showing similar UVW space motions, again utilizing the XHIP catalog.  We searched the XHIP catalog within a 5$\degr$ radius of HD 166191, requiring Hipparcos distances to be within 20 pc of the distance to HD 166191.  We then vetted for any stars within 3$\sigma$ of our measured UVW for HD 166191.  Only one star, HD 163296, passed all criteria (d$\sim$119 pc, U = -0.9 $\pm$ 3.3, V = -22.9 $\pm$ 2.2, and W = -7.4 $\pm$ 0.8).  HD 163296 is a well-studied Herbig Ae star hosting a substantial circumstellar disk with a wealth of silicate features and an estimated age of 5$^{+0.3}_{-0.6}$ Myr \citep{mon09}. This star has been proposed previously to be an outlying ScoCen member \citep{sit08}.  While HD 166191 and HD 163296 share similar galactic locations and space motions, HD 163296 show signs of active accretion (H$\alpha$ in emission - \citealt{silaj10}) and is detected in x-rays \citep{gun09}, indicating it may be younger than HD 166191.  However, it is possible that HD 166191 and HD 163296 are related outlying members of the ScoCen association.

\subsection{Infrared Excess/Silicate Feature}

While HD 166191 is mentioned in \cite{oud92}, \cite{fuj13}, and \cite{ken13} as an excess star, the full extent of its IR-excess and analysis of the unique shape of its SED has not yet been performed.  A planetary or brown dwarf companion could not produce the large IR luminosities seen in Figure 1.  We also rule out the possibility that HD 166191 has a transitional disk based on our age estimate in section 4.2, as transitional disks are dispersed by 10 Myr (\citealt{zuck95}, \citealt{carp06}, \citealt{sic06}, \citealt{cur09}).  In addition, the T-ReCs imaging rules out the possibility that the excess infrared emission is produced from interstellar dust grains creating a Pleiades-like effect, as any such emission would have been resolved (e.g. 29 Per - \citealt{zuck12}).  All available evidence indicates that the excess infrared emission emanates from a single source with a dusty disk.  We estimate the fraction of the stellar luminosity reradiated by HD 166191's surrounding dust ($\tau$ $\tbond$ L$_{IR}$/L$_{bol}$ $\sim$10\%) by integrating under the stellar SED and dust excess SED curves that appear in Figure 1.  The contribution from the hot component, which includes the 10$\mu$m feature, is $\sim$6\% L$_{bol}$, while the other $\sim$4\% L$_{bol}$ can be attributed to the cooler, more distant dust component. A detailed mineralogical study will require a mid-IR spectrum.  We do note though, that the peak of the solid-state emission feature ($\sim$10.3 $\mu$m) is indicative of silicates rather than silica, where the latter is typically associated with hypervelocity collisions of large objects \citep{lisse09}, implying that the warm dust excess has likely been produced via planetary collisions at velocities less than 5 km s$^{-1}$ or by the gradual grinding of asteroids.   

While the two blackbody dust fit model may not be a true interpretation of the dusty nature of this system, we can use this simple model to estimate some fundamental disk parameters. HD 166191 shows evidence of substantial quantities of both warm and cold orbiting dust particles.  The substantial amounts of warm dust orbiting HD 166191 are suggestive of ongoing collisional formation of rocky, terrestrial-like planetesimals (\citealt{rhee08}, \citealt{melis10} and references therein).  Assuming the warm dust grains around HD 166191 radiate like blackbodies, they lie at a distance of $\sim$0.3 AU.  Just as with the extremely dusty V488 Per \citep{zuck12}, this distance from the central star is comparable to that of many transiting planets discovered by NASA's {\it Kepler} satellite.  \cite{han12} propose a mechanism by which planets form at small distances from their host star without undergoing radial migration.  HD 166191 (and V488 Per) can be examples of this type of in situ planet formation.  Radiative blowout considerations can constrain the smallest allowable size of orbiting dust particles, which for HD 166191 is $\sim$0.3 $\mu$m.  We note that this size assumes no substantial amount of gas and, if the system is as young as 6 Myr, then this may not be a good assumption.  From equation (5) of \cite{rhee08}, the minimum dust mass for the hot dust component ($\tau$ $\sim$6\%) is 5 $\times$ 10$^{20}$ to 6 $\times$ 10$^{23}$ g for the same range of orbital radii and grain sizes mentioned in section 4.2, equivalent to the mass of a R = 40 to 400 km, 2.5 g cm$^{-3}$ rocky body.

\section{Conclusions}

We report an in-depth study of the age, origin, and IR-excess of the F8-type star HD 166191.  Based on theoretical evolutionary isochrones and optical spectroscopic observations, we estimate an age of 10-100 Myr for this star.  Mid- and far-IR photometry indicates substantial amounts of warm and hot dust, as well as a strong solid-state emission feature at $\sim$10 $\mu$m. The dust encircling HD 166191 is likely due to ongoing planet formation.  The fractional infrared luminosity of the HD 166191 system (L$_{IR}$/L$_{bol}$ $\sim$10\%) is among the largest known, along with V488 Per (L$_{IR}$/L$_{bol}$ $\sim$16\%; \citealt{zuck12}) and TYC 8241-2652-1 before its dust disappeared (L$_{IR}$/L$_{bol}$ $\sim$11\%; \citealt{melis12}).  

\acknowledgments

Based on observations obtained at the Gemini Observatory, which is operated by the Association of Universities for Research in Astronomy, Inc., under a cooperative agreement with the NSF on behalf of the Gemini partnership: the National Science Foundation (United States), the National Research Council (Canada), CONICYT (Chile), the Australian Research Council (Australia), Minist\'{e}rio da Ci\^{e}ncia, Tecnologia e Inova\c{c}\~{a}o (Brazil) and Ministerio de Ciencia, Tecnolog\'{i}a e Innovaci\'{o}n Productiva (Argentina). (Some of) The data presented herein were obtained at the W.M. Keck Observatory, which is operated as a scientific partnership among the California Institute of Technology, the University of California and the National Aeronautics and Space Administration. The Observatory was made possible by the generous financial support of the W.M. Keck Foundation.  This research has made use of the SIMBAD database and VizieR catalog access tool, operated at CDS, Strasbourg, France.  This publication makes use of data products from the Two Micron All Sky Survey, which is a joint project of the University of Massachusetts and the Infrared Processing and Analysis Center/California Institute of Technology, funded by the National Aeronautics and Space Administration and the National Science Foundation, and the {\it Wide-field Infrared Survey Explorer}, which is a joint project of the University of California, Los Angeles, and the Jet Propulsion Laboratory/California Institute of Technology, funded by the National Aeronautics and Space Administration.  This work is based [in part] on observations made with the Spitzer Space Telescope, which is operated by the Jet Propulsion Laboratory, California Institute of Technology under a contract with NASA. This research made use of data products from the Midcourse Space Experiment. Processing of the data was funded by the Ballistic Missile Defense Organization with additional support from NASA Office of Space Science. This research has made use of the NASA/ IPAC Infrared Science Archive, which is operated by the Jet Propulsion Laboratory, California Institute of Technology, under contract with the National Aeronautics and Space Administration.  The research was partially supported by a NASA grant to UCLA.  C.M. acknowledges support from the National Science Foundation under award No. AST-1003318.  A portion of this work was supported by the National Science Foundation under grant Nos. AST-1203023.

\begin{deluxetable}{lccc}
\tablecaption{HD 166191 Photometry\tablenotemark{a}}
\tablewidth{0pt}
\tablehead{
\colhead{Parameter} & \colhead{Wavelength ($\mu$m)} & \colhead{Flux (Jy)} & \colhead{Ref.}}
\startdata
B & 0.44 & 1.122 $\pm$ 0.021 & 1\\
V & 0.55 & 1.646 $\pm$ 0.023 & 1\\
J & 1.25 & 1.939 $\pm$ 0.038 & 2\\
H & 1.65 & 1.527 $\pm$ 0.060 & 2\\
K$_S$ & 2.20 & 1.100 $\pm$ 0.018 & 2\\
W1 & 3.35 & 0.780 $\pm$ 0.026 & 3\\
W2 & 4.60 & 0.874 $\pm$ 0.020 & 3\\
W3 & 11.56 & 1.978 $\pm$ 0.026 & 3\\
W4 & 22.09 & 3.241 $\pm$ 0.045 & 3\\
IRAS & 12.0 & 2.35 $\pm$ 0.752 & 4\\
IRAS & 25.0 & 3.80 $\pm$ 0.570 & 4\\
MSX A & 8.28 &  2.028 $\pm$ 0.083 & 5\\
MSX C & 12.13 & 1.863 $\pm$ 0.117 & 5\\
MSX D & 14.65 & 1.953 $\pm$ 0.129 & 5\\
MSX E & 21.34 & 3.166 $\pm$ 0.212 & 5\\
AKARI & 18 & 2.466 $\pm$ 0.0246 & 6\\
IRAC & 3.55 & 0.738 $\pm$ 0.061 & 7\\
IRAC & 4.49 & 0.674 $\pm$ 0.033 & 7\\
IRAC & 5.73 & 0.872 $\pm$ 0.025 & 7\\
IRAC & 7.87 & 1.621 $\pm$ 0.042 & 7\\
MIPS & 24 & 2.74 $\pm$ 0.01 & 8\\
MIPS & 70 & 1.73 $\pm$ 0.03 & 8\\
T-ReCS Si-1\tablenotemark{\dagger} & 7.73 & 1.51 $\pm$ 0.15 & 8\\
T-ReCS Si-2\tablenotemark{\dagger} & 8.74 & 2.44 $\pm$ 0.24 & 8\\
T-ReCS Si-3\tablenotemark{\dagger} & 9.69 & 2.89 $\pm$ 0.29 & 8\\
T-ReCS Si-4\tablenotemark{\dagger} & 10.38 & 3.22 $\pm$ 0.32 & 8\\
T-ReCS Si-5\tablenotemark{\dagger} & 11.66 & 2.71 $\pm$ 0.27 & 8\\
T-ReCS Si-6\tablenotemark{\dagger} & 12.33 & 2.00 $\pm$ 0.20 & 8\\
\enddata
\tablenotetext{a}{Not color-corrected.}
\tablenotetext{\dagger}{Bandwidths for T-ReCS silicate filters 1 through 6 are 
0.7, 0.9, 1.0, 1.0, 1.0, and 1.2 $\mu$m, respectively.}
\\
References: (1) Johnson magnitudes converted from Tycho 2 \citep{hog00}; (2) 2MASS \citep{cut03}; (3) WISE \citep{cut12}; (4) IRAS \citep{beich88}; (5) MSX \citep{price01}; (6) AKARI \citep{ish10}; (7) GLIMPSE \citep{ben03}; (8) This work
\end{deluxetable}

\begin{deluxetable}{lccc}
\tablecaption{HD 166191 Properties}
\tablewidth{0pt}
\tablehead{
\colhead{Parameter} & \colhead{Value} & \colhead{Ref.}}
\startdata
$\alpha$ (J2000) & 18:10:30.33 & 1\\ 
$\delta$ (J2000) & -23:34:00.3 & 1\\ 
$l$ ($\degr$) & 7.4400 & 1\\
$b$ ($\degr$)& -2.1427 & 1\\
$\mu$$_{\alpha}$ & -3.7 $\pm$ 1.6  mas yr$^{-1}$  & 2\\
$\mu$$_{\delta}$ & -40.2 $\pm$ 1.5  mas yr$^{-1}$ & 2\\
Parallax & 8.39 $\pm$ 1.16 mas & 3\\
Spectral Type\tablenotemark{a} & F8 $\pm$ 1 & 4\\
E(B-V)\tablenotemark{b} & 0.09 & 4\\
T$_{eff}$\tablenotemark{b} & $\sim$6300 K & 4\\
log $g$\tablenotemark{b} & 3.9 & 4\\
$[$M/H$]$\tablenotemark{b} & -0.25 & 4\\
RV & -8.1 $\pm$ 1.3 km s$^{-1}$ & 4\\
UVW\tablenotemark{c} & (-5.6 $\pm$ 1.3, -21.9 $\pm$ 3.0, -8.7 $\pm$ 1.5) (km s$^{-1}$)& 4\\
Li EW & 120 $\pm$ 5 m\AA & 4\\
H$\alpha$ EW & 2.7 $\pm$ 1 \AA & 4\\
L$_{IR}$/L$_{bol}$ & 0.10 & 4\\
\enddata
\tablenotetext{a}{Spectral type is determined by measuring numerous line depth ratios and utilizing the analytical expressions from \cite{kov03}.}
\tablenotetext{b}{E(B-V), T$_{eff}$, log $g$, and [M/H] are determined by fitting model atmosphere fluxes to the WiFeS spectrum.}
\tablenotetext{c}{UVW space motions are calculated using the measured radial velocity. $UVW$ are defined with respect to the Sun. U is positive toward the Galactic center, V is positive in the direction of Galactic rotation, and W is positive toward the north Galactic pole.}  
\\
References: (1) 2MASS catalog \citep{cut03}; (2) Tycho 2 catalog \citep{hog00}; (3) Hipparcos catalog \citep{van07}; (4) This work.
 \end{deluxetable}

\begin{figure}
\plotone{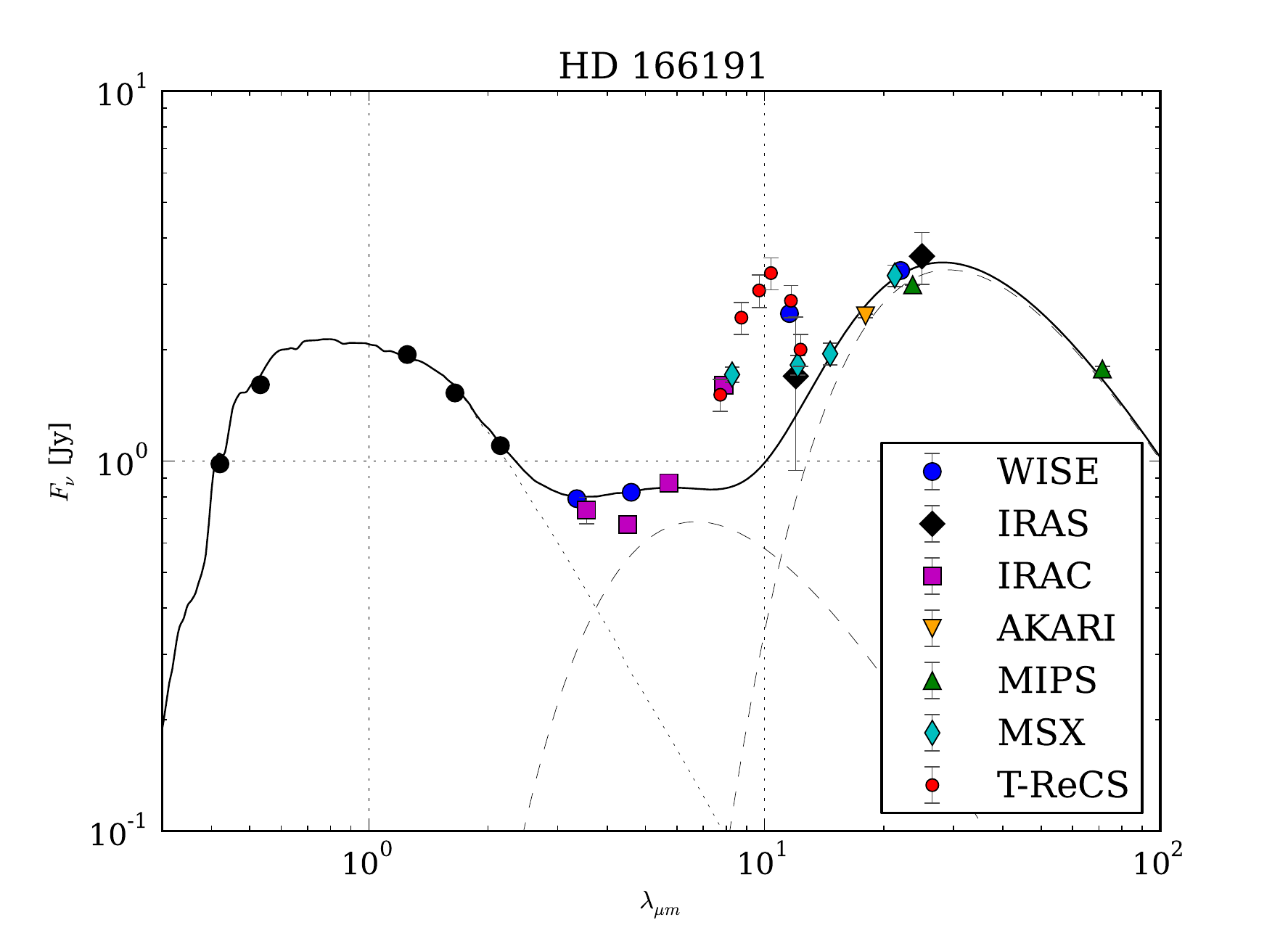}
\caption{The spectral energy distribution of HD 166191.  Black circles represent B, V, J, H, and K photometry from Tycho-2 and 2MASS, ordered from left to right.  All other symbols are specified in the figure legend.  In some cases (IRAC, WISE), the error bars are smaller than the symbol size.  The dotted line is the model fit to the Tycho-2 and 2MASS data.  The dashed curves are blackbody fits to the excess emission with temperatures of 760 and 175 K.  The solid line is the combined stellar and blackbody dust fit.  Color corrections have been applied to all broadband photometric measurements greater than 3 $\mu$m.
}
\end{figure}

\begin{figure}
\plotone{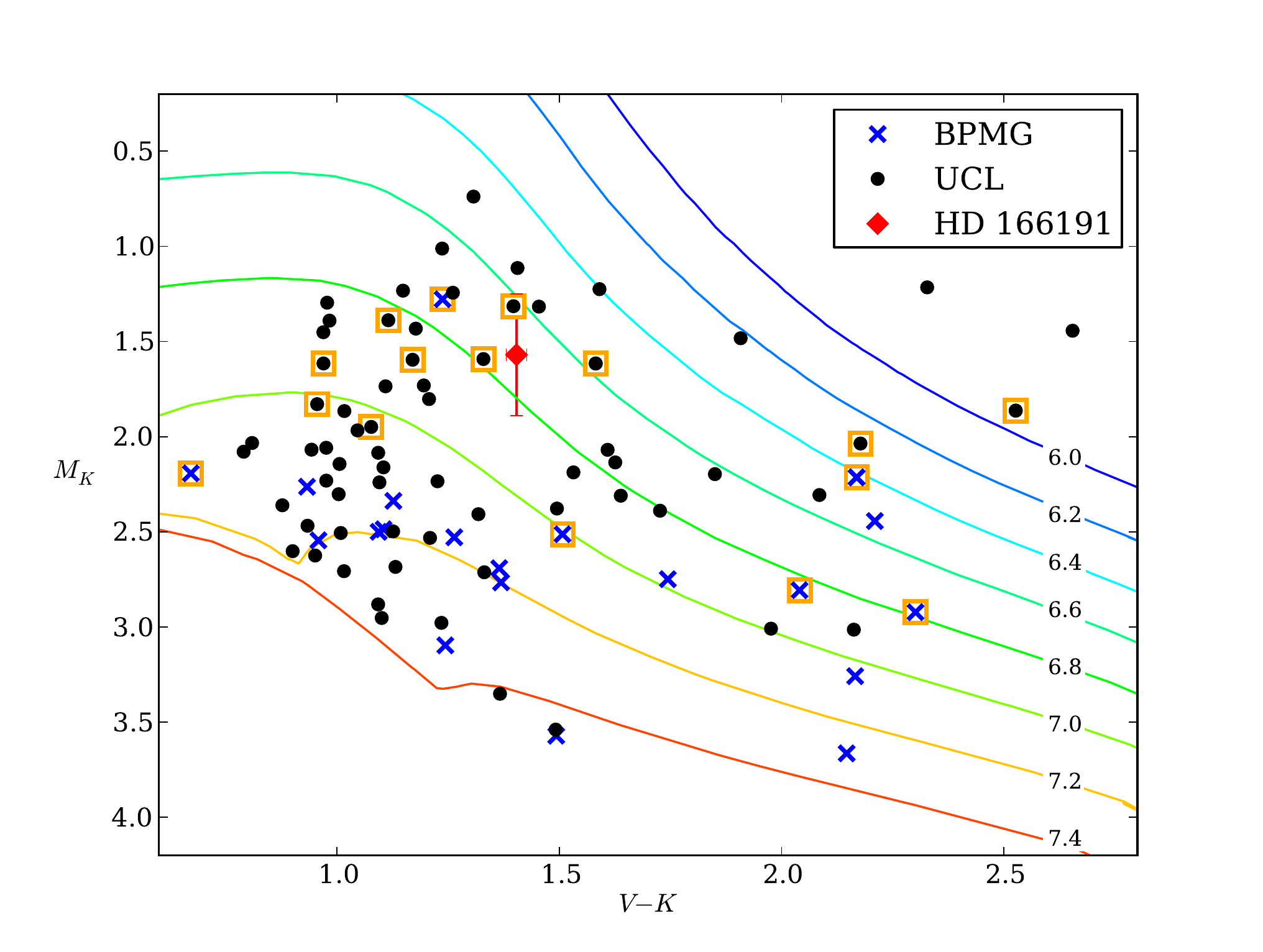}
\caption{A comparison of the position of HD 166191 on the color-magnitude diagram with theoretical isochrones \citep{bre12}.  Isochrones are labeled with log(t/yr) values.  The black circles are accepted Upper-Centaurus-Lupus (UCL) Hipparcos members from \cite{chen11} while the blue crosses are members of the Beta Pictoris Moving Group.  Orange boxes indicate known binaries.  The error bar for HD 166191 represents the combined photometric and parallax uncertainties.}
\end{figure}

\begin{figure}
\plotone{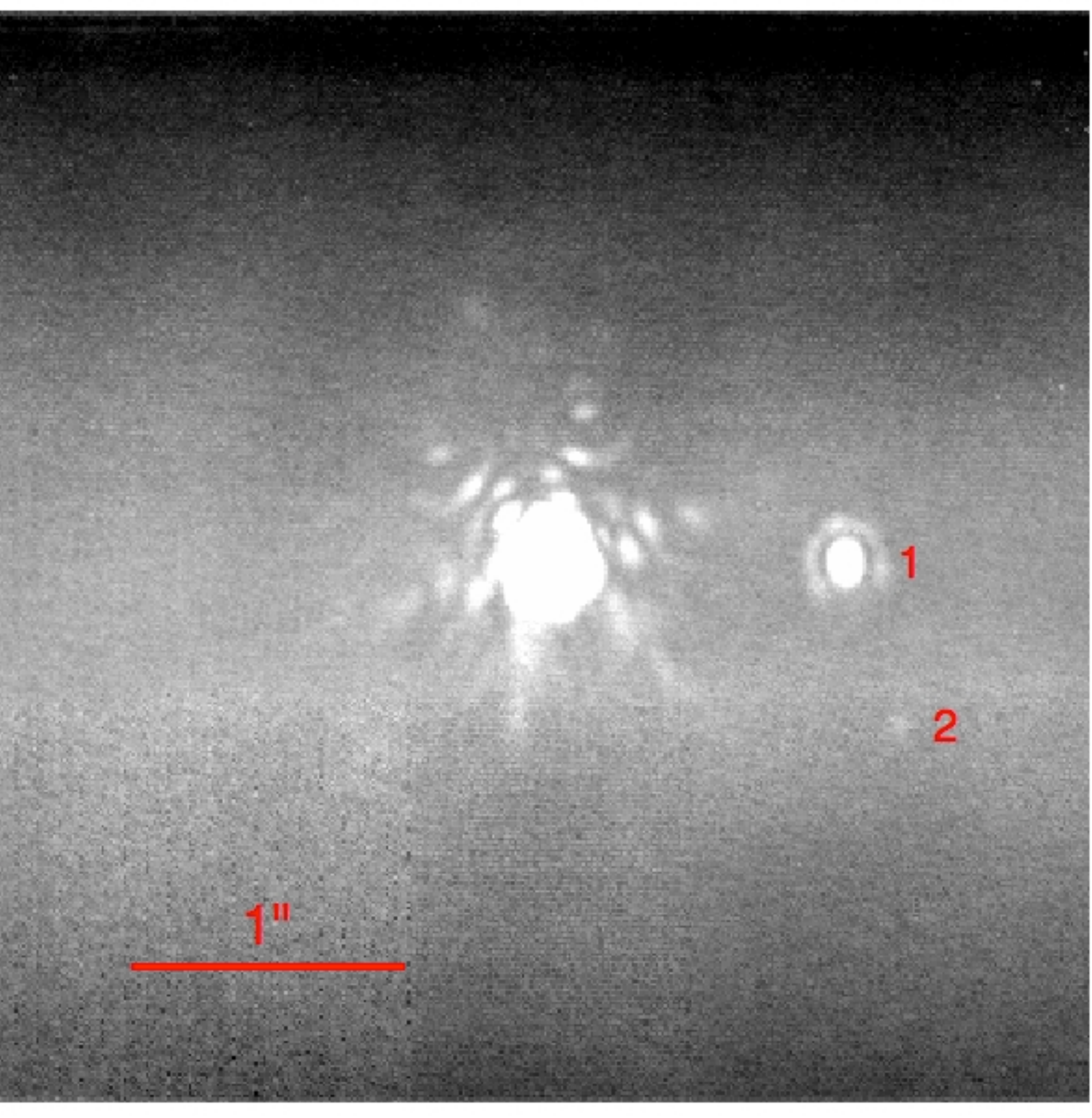}
\caption{Lp image of HD 166191 obtained with the NIRC2 camera.  Additional components (labeled 1 and 2) are discussed in section 4.2.}
\end{figure}


\begin{thebibliography}{}
\bibitem[Anderson \& Francis(2012)]{and12} Anderson, E., \& Francis, C., 2012, Astron. Lett, 38, 331
\bibitem[Beichman et al.(1988)]{beich88} Beichman, C. A., Neugebauer, G., Habing, H. J., Clegg, P. E., \& Chester, T. J., 1988, Infrared Astronomical Satellite (IRAS) Catalogs and Atlases, Vol. 1, catalog and explanatory supplement. 
\bibitem[Benjamin et al.(2003)]{ben03} Benjamin, R. A., Churchwell, E., Babler, B. L., et al., 2003, \pasp, 121, 76
\bibitem[Bressan et al.(2012)]{bre12} Bressan, A., Marigo, P., Girardi, L., et al., 2012, \mnras, 427, 127
\bibitem[Bryden et al.(2006)]{bry06} Bryden, G., Beichman, C. A., Trilling, D. E., et al., 2006, \apj, 636, 1098
\bibitem[Bryden et al.(2009)]{bry09} Bryden, G., Beichman, C. A., Carpenter, J. M., et al., 2009, \apj, 705, 1226
\bibitem[Carey et al.(2009)]{car09} Carey, S. J., Noriega-Crespo, A., Mizuno, D. R., et al., 2009, \pasp, 115, 953
\bibitem[Carpenter et al.(2006)]{carp06} Carpenter, J. M., Mamajek, E. E., Hillenbrand, L. A., \& Meyer, M. R., 2006, \apj, 651, 49
\bibitem[Chen \& Jura(2001)]{chen01} Chen, C. H., \& Jura, M., 2001, \apjl, 560, L171
\bibitem[Chen et al.(2005)]{chen05} Chen, C. H., Jura, M., Gordon, K. D., \& Blaylock, M., 2005, \apj, 623, 493
\bibitem[Chen et al.(2006)]{chen06} Chen, C. H., Sargent, B. A., Bohac, C., et al., 2006, \apjs, 166, 351
\bibitem[Chen et al.(2011)]{chen11} Chen, C. H., Mamajek, E. E., Bitner, M. A., et al., 2011, \apj, 738, 122
\bibitem[Currie et al.(2007)]{cur07} Currie, T., Kenyon, S. J., Rieke, G., Balog, Z., \& Bromley, B. C., 2007, \apj, 663, L105
\bibitem[Currie et al.(2008a)]{cur08a} Currie, T., Kenyon, S. J., Balog, Z., et al., 2008a, \apj, 672, 558
\bibitem[Currie et al.(2008b)]{cur08b} Currie, T., Plavchan, P., \& Kenyon, S. J., 2008b, \apj, 688, 597
\bibitem[Currie et al.(2009)]{cur09} Currie, T., Lada, C. J., Plavchan, P., et al., 2009, \apj, 698, 1
\bibitem[Cutri et al.(2003)]{cut03} Cutri, R. M., Skrutskie, M. F., van Dyk, S., et al. 2003, VizieR Online Data Catalog, 2246, 0
\bibitem[Cutri et al.(2012)]{cut12} Cutri, R. M., et al. 2012, VizieR Online Data Catalog, 2311, 0
\bibitem[de Zeeuw et al.(1999)]{dez99} de Zeeuw, P. T., Hoogerwerf, R., de Bruijne, J. H. J., Brown, A. G. A., \& Blaaw, A., 1999, \aj, 117, 354
\bibitem[Fazio et al.(2004)]{faz04} Fazio, G., Hora, J. L., Allen, L. E., et al., 2004, \apjs, 154, 10
\bibitem[Fujiwara et al.(2013)]{fuj13} Fujiwara, H., Ishihara, D., Onaka, T., et al., 2013, \aap, 550, 45
\bibitem[G\"{u}nther \& Schmitt(2009)]{gun09} G\"{u}nther, H. M. \& Schmitt, J. H. M. M., 2009, \aap, 494, 1041
\bibitem[Hansen \& Murray(2012)]{han12} Hansen, B. \& Murray, N., 2012, \apj, 751, 158
\bibitem[Hoaglin, Mostellar, \& Turkey(1983)]{hoag83} Hoaglin, D. C., Mostellar F., \& Turkey, F. J., 1983, Understanding Robust and Exploratory Data Analysis, Wiley, New York
\bibitem[H\o g et al.(2000)]{hog00} H\o g, E., Fabricius, C., Makarov, V. V., et al., 2000, \aap, 355, 27
\bibitem[Ishihara et al.(2010)]{ish10} Ishihara, D., Onaka, T., Kataza, H., et al., 2010, \aap, 514, A1
\bibitem[Johnson et al.(2012)]{john12} Johnson, B. C., Lisse, C. M., Chen, C. H., et al., 2012, \apj, 761, 45
\bibitem[Kennedy \& Wyatt(2013)]{ken13} Kennedy, G. M., \& Wyatt, M. C., 2013, preprint (arXiv: 1305.6607)
\bibitem[Kenyon \& Bromley(2008)]{ken08} Kenyon, S. J., \& Bromley, B. C., 2008, \apjs, 179, 451
\bibitem[Kovtyukh et al.(2003)]{kov03} Kovtyukh, V. V., Soubiran, C., Belik, S. I., \& Gorlova, N. I., 2003, \aap, 411, 559
\bibitem[Lawrence et al.(2007)]{law07} Lawrence, A., Warren, S. J., Almaini, O., et al., 2007, \mnras, 379, 1599
\bibitem[Lisse et al.(2002)]{lisse02} Lisse, C. M., Schultz, A., Fernandez, Y., et al., 2002, \apj, 570, 779
\bibitem[Lisse et al.(2009)]{lisse09} Lisse, C. M., Chen, C. H., Wyatt, M. C., et al., 2009, \apj, 701, 2019
\bibitem[Lucas et al.(2008)]{luc08} Lucas, P. W., Hoare, M. G., Longmore, A., et al., 2008, \mnras, 391, 136
\bibitem[Makarov \& Marleau(2005)]{mak05} Makarov, D., \& Marleau, F., 2005, \pasp, 117, 1113
\bibitem[Melis et al.(2010)]{melis10} Melis, C., Zuckerman, B., Rhee, J. H., \& Song, I., 2010, \apj, 717, 57
\bibitem[Melis et al.(2012)]{melis12} Melis, C., Zuckerman, B., Rhee, J. H., et al., 2012, \nat, 487, 74
\bibitem[Meng et al.(2012)]{meng12} Meng, H., Rieke, G., Su, K., et al., 2012, \apj, 751, L17
\bibitem[Montesinos et al.(2009)]{mon09} Montesinos, B., Eiroa, C., Mora, A., \& Merin, B., 2009, \aap, 495, 901
\bibitem[Morales et al.(2011)]{mor11} Morales, F. Y., Rieke, G. H., Werner, M. W., et al., 2011, \apj, 730, 29
\bibitem[Ortega et al.(2002)]{ort02} Ortega, V. G., de la Reza, R., Jilinski, E., \& Bazzanella, B., 2002, \apj, 575, 57
\bibitem[Oudmaijer et al.(1992)]{oud92} Oudmaijer, R., van der Veen, W., Waters, L., et al., 1992, \aaps, 96, 625
\bibitem[Owen, Ercolano, \& Clarke(2011)]{owen11} Owen, J. E., Ercolano, B., \& Clarke, C. J., 2011, \mnras, 412, 13
\bibitem[Price et al.(2001)]{price01} Price, S. D., Egan, M. P., Carey, S. J., Mizuno, D. R., \& Kuchar, T. A., 2001, \aj, 121, 2819
\bibitem[Rhee et al.(2007)]{rhee07} Rhee, J. H., Song, I., Zuckerman, B., \& McElwain, M., 2007, \apj, 660, 1556
\bibitem[Rhee et al.(2008)]{rhee08} Rhee, J. H., Song, I., \& Zuckerman, B., 2008, \apj, 675, 777
\bibitem[Rieke et al.(2004)]{rieke04} Rieke, G. H., Young, E. T., Engelbracht, C. W., et al., 2004, \apjs, 154, 25
\bibitem[Sicilia-Aguilar et al.(2006)]{sic06} Sicilia-Aguilar, A., Hartmann, L., Calvet, N., et al., 2006, \apj, 638, 897
\bibitem[Silaj et al.(2010)]{silaj10} Silaj, J., Jones, C. E., Tycner, C., Sigut, T., \& Smith, A. D., 2010, \apjs, 187, 228
\bibitem[Sitko et al.(2008)]{sit08} Sitko, M. L., Carpenter, W. J., Kimes, R. L., et al., 2008, \apj, 678, 1070
\bibitem[Smith et al.(2008)]{smith08} Smith, R., Wyatt, M. C., \& Dent, W. R. F., 2008, \aap, 485, 897
\bibitem[Song et al.(2005)]{song05} Song, I., Zuckerman, B., Weinberger, A. J., \& Becklin, E. E., 2005, \nat, 436, 363
\bibitem[Telesco et al.(1998)]{tel98} Telesco, C. M., Pina, R. K., Hanna, K. T., et al., 1998, Proc. SPIE, 3354,534
\bibitem[van Leeuwen(2007)]{van07} van Leeuwen, F., 2007, \aap, 474, 653
\bibitem[Watson et al.(2009)]{wat09} Watson, M. G., Schroder, A. C., Fyfe, D., et al., 2009, \aap, 493, 339
\bibitem[Wright et al.(2010)]{wri10} Wright, E. L., Eisenhardt, P., Mainzer, A., et al., 2010, \aj, 140, 1868
\bibitem[Wyatt(2008)]{wyatt08} Wyatt, M. C., 2008, \araa, 46, 339
\bibitem[Zuckerman et al.(1995)]{zuck95} Zuckerman, B., Forveille, T., \& Kastner, J. H., 1995, \nat, 373, 494
\bibitem[Zuckerman \& Song(2004)]{zuck04} Zuckerman, B. \& Song, I., 2004, \araa, 42, 685
\bibitem[Zuckerman et al.(2012)]{zuck12} Zuckerman, B., Melis, C., Rhee, J. H., Schneider, A. \& Song, I., 2012, \apj, 752, 58
\end{thebibliography}
\end{document}